\documentclass[preprint,aps,amsmath,nofootinbib,tightenlines]{revtex4}
\usepackage{graphicx}
\usepackage{bm}
\usepackage{epsfig}

\newif\ifpdf
\ifx\pdfoutput\undefined
\pdffalse 
\else
\pdfoutput=1 
\pdftrue
\fi

\def\itbf#1{\mbox{\boldmath $#1$}}

\def\Dsl{\hbox{/\kern-.6000em D}} 

\def\dsl{\,\raise.15ex\hbox{/}\mkern-13.5mu D}

\def\psip#1{\psi_{\mathbf{#1}}}
\def\chip#1{\chi_{\mathbf{#1}}}

\def\sbsigma{\mbox{\scriptsize\boldmath $\sigma$}}

\def\ltap{\ \raise.3ex\hbox{$<$\kern-.75em\lower1ex\hbox{$\sim$}}\ }
\def\gtap{\ \raise.3ex\hbox{$>$\kern-.75em\lower1ex\hbox{$\sim$}}\ }
\def\OMIT#1{}

\def\lsim{\mathrel{\raise.3ex\hbox{$<$\kern-.75em\lower1ex\hbox{$\sim$}}}}
\def\gsim{\mathrel{\raise.3ex\hbox{$>$\kern-.75em\lower1ex\hbox{$\sim$}}}}
\def\sitbf#1{\mbox{\scriptsize\boldmath $#1$}}

\newcommand{\nn}{\nonumber}

\newcommand{\bmk}{\mathbf k}

\newcommand{\bbmp}{\mbox{\scriptsize\boldmath $p$}}

\catcode`\@=11
\def\slash{\mathpalette\make@slash}
\def\make@slash#1#2{\setbox\z@\hbox{$#1#2$}%
  \hbox to 0pt{\hss$#1/$\hss\kern-\wd0}\box0}
\catcode`\@=12 

\begin{document}
\ifpdf
\DeclareGraphicsExtensions{.pdf, .jpg}
\else
\DeclareGraphicsExtensions{.eps, .jpg}
\fi


\preprint{ \vbox{ \hbox{MPP-2005-30} 
}}

\title{\phantom{x}\vspace{0.5cm} 
The Large Higgs Energy Region in Higgs Associated Top Pair Production
at the Linear Collider
\vspace{1.0cm} }

\author{Cailin Farrell and Andr\'e~H.~Hoang\vspace{0.5cm}}
\affiliation{Max-Planck-Institut f\"ur Physik\\
(Werner-Heisenberg-Institut), \\
F\"ohringer Ring 6,\\
80805 M\"unchen, Germany\vspace{1cm}
\footnote{Electronic address: ahoang@mppmu.mpg.de, farrell@mppmu.mpg.de}\vspace{1cm}}


\begin{abstract}
\vspace{0.5cm}
\setlength\baselineskip{18pt}
The process $e^+e^-\to t\bar t H$ is considered in the kinematic
end point region where the Higgs energy is close to its maximal energy.
In perturbative QCD, using the loop expansion, the amplitudes are plagued by
Coulomb singularities that need to be resummed. We show that the QCD
dynamics in this end point region is governed by nonrelativistic heavy
quarkonium dynamics, and we use a nonrelativistic effective theory to
compute the Higgs energy distribution at leading and 
next-to-leading-logarithmic approximation in the nonrelativistic
expansion. Updated umbers for the total cross section including the
summations in the Higgs energy end point region are presented. 
\end{abstract}
\maketitle


\newpage

%
%
%
\section{Introduction}
\label{sectionintroduction}

It is one of the most important tasks of future collider experiments to unravel
in detail
the mechanism of mass generation and electroweak symmetry breaking. Within the
Standard Model of particle physics (SM) electroweak symmetry breaking
is achieved   
by the Higgs mechanism which postulates the existence of an electrically
neutral scalar field that interacts with all SM particles carrying non-zero
hypercharge and weak isospin.  The particle masses are then generated by the
Higgs field vacuum expectation value $V=(\sqrt{2}\,G_F)^{1/2}\approx 246$~GeV,
$G_F$ being the Fermi constant, that arises through the Higgs self
interactions. The mechanism also predicts that the Higgs field can be manifest
as a massive, elementary, scalar Bose particle that can be produced in
collider experiments. The mass of the Higgs boson is expected to lie between
the current lower experimental limit of $114.4$~GeV~\cite{LEPlimits} and about
$1$~TeV. Current analyses of electroweak precision observables yield a $95\%$
CL upper bound of $260$~GeV for the Higgs boson
mass~\cite{MHupperlimit}. While a Higgs boson with a mass smaller than $1$~TeV
can be found at the LHC, precise and model-independent measurements of its
quantum numbers and couplings can be gained from the $e^+e^-$ Linear
Collider~\cite{TESLATDR,ALCphysics,ACFALCphysics}.  

An important prediction of the Higgs mechanism is that its Yukawa couplings to
quarks $\lambda_q$ are related to the quark masses by   $m_q=\lambda_q
V$. This makes the measurement of the Yukawa coupling to the top quark
particularly important since it can be expected that it will be determined with
the highest precision among the Yukawa couplings, either through direct or
indirect methods. At the $e^+e^-$ Linear Collider the top quark Yukawa coupling
can be measured from top quark pair production associated with a
Higgs boson, $e^+e^-\to t\bar t H$. This process is particulary suited for
a light Higgs boson since the cross section is then large and the rate is
dominated by the amplitudes describing Higgs radiation off the top or the
antitop quark. Assuming an experimental precision at the percent level, QCD
and electroweak radiative corrections need to be accounted for in the
theoretical predictions. 

Approximating the top quark and the Higgs boson as stable particles the Born
cross section was already determined some time ago in
Refs.~\cite{Borneetth}. For the 
${\cal O}(\alpha_s)$ QCD one-loop corrections a number of references in various 
approximations exist. An analysis for large energy and small Higgs
masses can be found in Ref.~\cite{Dawson1}, while a full computation
of the numerically dominant 
virtual photon exchange contributions, where the Higgs is radiated exclusively
from the top and antitop quark, was presented in Ref.~\cite{Dawson2}. Finally,
the full set of QCD corrections was given in Ref.~\cite{Dittmaier1}. On the
other hand, the 
full set of one-loop electroweak corrections was obtained in
Refs.~\cite{Belanger1,Denner1} and also in Refs.~\cite{You1}. 
In Ref.~\cite{Denner1} a detailed analysis of various differential
distributions of 
the cross section $\sigma(e^+e^-\to t\bar t H)$ was given.

%
%
\begin{figure}[t] 
\begin{center}
 \leavevmode
 \epsfxsize=9cm
 \leavevmode
 \epsffile[140 300 430 495]{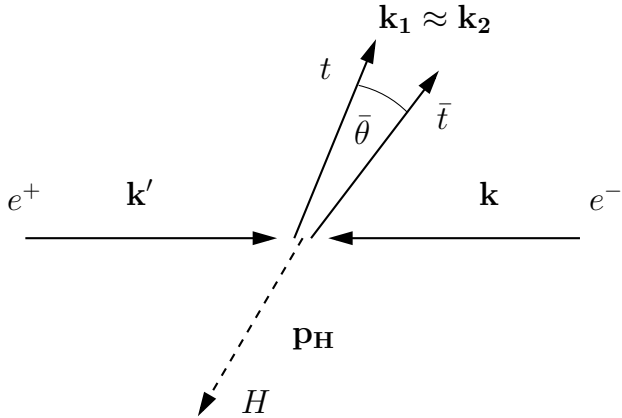}
 \vskip  0.0cm
 \caption{
Typical constellation of momenta for the process $e^+e^-\to t\bar t H$ in the
large Higgs energy end point region. 
 \label{fig1} }
\end{center}
\end{figure}
A particularly
interesting kinematical phase space region is where the energy of the Higgs
boson is large and close to its kinematic end point.  In this end point region
the $t\bar t$ pair is forced to become collinear and to fly opposite
to the Higgs direction in order to maximize the momentum necessary to balance
for the large Higgs momentum, see Fig.~\ref{fig1} for an illustration. The
general relation between the Higgs energy $E_H$ and the $t\bar t$ invariant mass
$Q^2=(k_1+k_2)^2$ reads
\begin{eqnarray}
E_H \, = \, \frac{1}{2\sqrt{s}}\,\left(s+m_H^2-Q^2\right) 
\,,
\end{eqnarray}  
where $\sqrt{s}$ is the c.m.\,energy and $m_H$ the Higgs boson mass. Thus for
large $E_H$ the $t\bar t$ invariant mass approaches $4m_t^2$ and the top quark
pair is nonrelativistic in its center-of-mass system. Since a light Higgs
boson having a mass below the $W^+W^-$ threshold is quite narrow\footnote{ 
For $m_H=115(150)$~GeV one finds $\Gamma_H=0.003(0.017)$~GeV~\cite{Hdecay}.}
and the hadronic Higgs decay final state factorizes to a very good
approximation, the strong interactions between the $t\bar t$ pair and the
hadronic Higgs final state can be neglected. Therefore, close to the Higgs
energy end point, the $t\bar t$ QCD dynamics is exclusively governed by
the nonrelativistic physics known from the process $e^+e^-\to t\bar t$ in the
$t\bar t$ threshold region at $\sqrt{s}\approx
2m_t$~\cite{TTBARreview}.   

In this regime the so-called Coulomb singularities $\propto (\alpha_s/v)^n$,
with $v=(1-4m_t^2/Q^2)^{1/2}$ being the top quark relative velocity in the
$t\bar t$ c.m.\,frame, arise in the amplitudes and render the perturbative
expansion in the number of loops inapplicable. This singularity structure is
most easily visible in the Higgs energy distribution, $d\sigma(e^+e^-\to t\bar
t H)/d E_H$. While the Born distribution approaches zero for 
$E_H\to E_H^{\rm max}$,
$d\sigma/d E_H\sim v$~\cite{Borneetth},
the ${\cal O}(\alpha_s)$ fixed-order perturbative corrections are proportional 
to $\alpha_s$ at the end point~\cite{Dawson2,Dittmaier1} and the ${\cal O}(\alpha_s^2)$
corrections even diverge like 
$\alpha_s^2/v$. In principle, the problem can be avoided by imposing a cut on
the Higgs energy (or the $t\bar t$ invariant mass $Q^2$), but such a measure
is unnecessary because there exists an 
elaborate technology from the threshold region in the process $e^+e^-\to t\bar
t$~\cite{TTBARreview}, that 
allows to sum the Coulomb singularities to all orders in $\alpha_s$ and
to carry out a simultaneous expansion in $\alpha_s$ and $v$ using a
nonrelativistic effective theory. Imposing a cut is also disadvantageous
since the nonrelativistic portion of the $t\bar t H$ phase space increases 
for smaller c.m.\,energies or larger Higgs masses. 
Due to the large top quark SM width
$\Gamma_t\approx 1.5$~GeV, which serves as an infrared cutoff, the
corresponding QCD effective theory computations can be carried out 
with perturbative methods for all Higgs energies in the 
end point region. The effective theory also allows for a systematic summation of
logarithmic terms $\propto (\alpha_s\ln v)^n$ to all orders in $\alpha_s$
using the velocity renormalization group~\cite{LMR}.  

In this paper we compute the Higgs energy distribution $d\sigma/d E_H$ in the
large Higgs energy end point region at leading-logarithmic (LL) and
next-to-leading-logarithmic (NLL) order in the nonrelativistic
expansion in the framework of "velocity" nonrelativistic QCD (vNRQCD)
using the conventions and notations of 
Refs.~\cite{LMR,amis,amis2,HoangStewartultra,hmst}.\footnote{
For an alternative approach, see Refs.~\cite{Brambillareview}.}
We neglect NLL order
effects coming from the top quark finite 
lifetime that were discussed recently in Ref.~\cite{HoangReisser1}. Our
analysis contributes to an improved understanding of uncertainties of 
higher order QCD corrections in the large Higgs energy region and is
phenomenologically important for the
total cross section  if the $e^+e^-$ c.m.\,\,energy is not too large.

The program of the paper is as follows: 
In Sec.~\ref{sectionEFT} the ingredients of the effective theory 
that are necessary for the calculation at hand are presented. Within the
effective theory framework the LL Higgs energy distribution in the
large Higgs energy end point is calculated explicitly as an
illustration in Sec.~\ref{sectionLL}. The computation of the
NLL Higgs energy distribution including the numerical determination of
the NLL matching conditions is outlined in
Sec.~\ref{sectionmatching}. A numerical 
analysis of the results is performed in Section~\ref{sectionanalysis}, and
Section~\ref{sectionconclusion} contains the conclusion.

\section{Effective Theory Ingredients} 
\label{sectionEFT}

In our calculations we employ vNRQCD (``velocity'' NRQCD), an effective theory
which describes the 
nonrelativistic dynamics of top quark pairs where the nonrelativistic
scales $m_t v$ (momentum) and $m_t v^2$ (energy) are larger than the
hadronization scale $\Lambda_{\rm QCD}$. It has become the standard
convention to call the momentum scale ``soft'' and the energy scale
``ultrasoft''. In the following we summarize the
ingredients necessary for the description of the nonrelativistic $t\bar t$
dynamics at the NLL order approximation in the $t\bar t$ c.m.\,frame. For
details on the conceptual aspects, concerning powercounting, the operator
structure of the effective theory action, and renormalization we refer to
Refs.~\cite{LMR,amis,amis2,HoangStewartultra,hmst}. 

The particle-antiparticle propagation is described by the terms in the
effective theory Lagrangian that are bilinear in the top quark and antitop quark
fields,
\begin{eqnarray} \label{Lke}
 {\mathcal L}(x) &=& \sum_{\bbmp}
   \psip{\bbmp}^\dagger(x)   \biggl\{ i \partial^0 - {\itbf{p}^2 \over 2 m_t}   
   + \frac{i}{2} \Gamma_t 
   - \delta m_t \biggr\} \psip{\bbmp}(x) 
+ (\psip{\bbmp}(x) \to\chip{\bbmp}(x))
\,,
\end{eqnarray}
where the fields $\psip{\bbmp}$ and $\chip{\bbmp}$ destroy top and antitop quarks with
soft three-momentum ${\itbf{p}}$ in the $t\bar t$ c.m.\,\,frame and $\Gamma_t$ is the
on-shell top quark decay width. The x-dependence of the fields describes
ultrasoft ($\sim m_t v^2$) fluctuations while the labels ${\itbf{p}}$ refer to
the soft ($\sim m_tv$) three-momentum of the quarks.
The term $\delta m_t$ is a residual mass term
that is specific to the top quark mass definition that is being
used. In order to avoid the pole mass 
renormalon problem~\cite{Aglietti1,Vcrenormalon} we employ the 1S mass
scheme~\cite{Hoangupsilon,HoangTeubnerdist} with 
\begin{eqnarray}
\delta m_t & = & m_{\rm 1S}\,\left\{\,
\Delta^{\mbox{\tiny LL}} + \Delta^{\mbox{\tiny NLL}}\,\right\}
\,,\label{mpolem1s} \\[2mm]
\Delta^{\mbox{\tiny LL}}(\nu) & = & \frac{a^2}{8}
\,,\\[2mm]
 \Delta^{\mbox{\tiny NLL}}(\nu) & = & \frac{a^3}{8\pi\, C_F}\, 
 \,\left[\, \beta_0\,\left( \ln\left(\frac{\nu}{a}\right) + 1 \right) 
+ \frac{a_1}{2}  \,\right]
\,,\\[2mm]
a & \equiv & -\frac{{\cal V}^{(s)}_c(\nu)}{4\pi}
\,,
\end{eqnarray} 
where $\beta_0=11/3 C_A-4/3 T n_f$ is the one-loop QCD beta function, 
$a_1= 31/9 C_A - 20/9 T n_f$ the coefficient of the one-loop correction to the
effective Coulomb potential and $C_A=3, C_F=4/3, T=1/2$ are SU(3) group
theoretical factors. For the number of light flavors we take $n_f=5$. The
parameter $\nu$ is the dimension-zero vNRQCD 
renormalization scaling parameter used to describe the correlated
running of soft and ultrasoft effects in the effective
theory governed by the velocity renormalization
group\,\cite{LMR}.\footnote{
The renormalization scales for soft and ultrasoft
fluctuations, $\mu_S$ and $\mu_U$, are correlated through the heavy
quark equation of motion, $\mu_U=\mu_S^2/m_t$. This correlation
is necessary for a consistent renormalization of the nonrelativistic
effective theory \cite{Hoang:2003ns}. The correlated running from the hard 
scale down to the soft and ultrasoft scales is described by the
dimensionless scaling parameter $\nu$ defined by $\mu_S=m_t\nu$ and
$\mu_U=m_t\nu^2$. Thus, $\nu=1$ corresponds to the hard matching scale,
and effective theory matrix elements are evaluated for $\nu\approx
0.2$, i.e. of order of the typical top quark velocity.  
}
The resulting
top/antitop propagator in the effective theory has the form
\begin{eqnarray}
  \frac{i}{p^0 - {\itbf{p}}^2/(2m_t) + i\Gamma_t/2 -\delta m_t} \,.
\label{toppropagator}
\end{eqnarray}

Up to NLL order the top-antitop quark pair interacts only through the effective
Coulomb potential~\cite{Fischler1,Billoire1},
\begin{eqnarray}
 \tilde V_c({\itbf{p}},{\itbf{q}}) 
 & = &
 \frac{{\cal{V}}_c^{(s)}(\nu)}{{\itbf{k}}^2}\, 
 -\,\frac{4\pi C_F\, \alpha_s(m_t\nu)}{{\itbf{k}}^2}\, \left\{\,
 \frac{\alpha_s(m_t\nu)}{4\pi}\,\left[\,
 -\beta_0\,\ln\Big(\frac{{\itbf{k}}^2}{m_t^2\nu^2}\Big) + a_1
 \,\right]\,\right\} 
\,,
\nonumber \\[2mm]
{\cal V}_c^{(s)}(\nu) & = & -4\pi C_F \alpha_s(m_t \nu)
\,,
 \label{VCoulomb}
\end{eqnarray}
where ${\itbf{k}}={\itbf{p}}-{\itbf{q}}$ is the momentum transfer. 
The term 
${\cal V}_c^{(s)}(\nu)$ is the color singlet Wilson coefficient of the
4-quark Coulomb potential operator. Here, $\nu=1$ corresponds to the hard
scale at which the effective theory is matched to the full theory, and
$\nu=v_0$, $v_0$ being of the order of the typical $t\bar t$ relative
velocity, is the scale where the matrix elements are computed. The evolution
of the Wilson coefficients from the matching scale down to the low-energy
scale sums logarithms of the velocity to all orders and
is governed by the velocity renormalization group equations~\cite{LMR}
which are determined from the anomalous dimensions of the effective theory. 

Top-antitop quark production in the nonrelativistic regime in the LL and NLL
approximation in a ${}^3S_1$ spin triplet or a ${}^1S_0$ spin singlet state
is described by the currents 
\begin{eqnarray}
  J^j_{1,{\sitbf{p}}} & = &
    \psi_{{\sitbf{p}}}^\dagger\, \sigma_j (i\sigma_2) \chi_{-{\sitbf{p}}}^*
   \,,
\nonumber\\[2mm] 
  J_{0,{\sitbf{p}}} & = &  \psi_{{\sitbf{p}}}^\dagger\,  (i\sigma_2)
    \chi_{-{\sitbf{p}}}^*
\,, 
\label{J1J0}
\end{eqnarray}
where $c_{1,j}(\nu)$ and $c_0(\nu)$ are the corresponding Wilson coefficients.
The currents do not run at LL order, but they have a non-trivial
anomalous dimension at NLL order from UV divergences in effective theory
two-loop vertex diagrams~\cite{LMR,HoangStewartultra,Pineda1}. The NLL running
of the Wilson coefficients 
reads
\begin{eqnarray}
c_{1,j}(\nu) & = & c_{1,j}(1)\,\exp\left(f(\nu,2) \right) \,,
\nonumber\\[4mm]
c_{0}(\nu) & = & c_{0}(1)\,\exp\left(f(\nu,0) \right)\,,
\label{currentWilson}
\end{eqnarray}
where
\begin{eqnarray}
f(\nu,{\bf S^2}) & = & {} + a_2\,\pi\alpha_s(m_t)\,\left(1-z\right) 
  {}+ a_3\, \pi \alpha_s(m_t) \ln(z) 
\nn\\[4mm] &&
 {}+ a_4\, \pi\alpha_s(m_t) \Big[\,1- z^{1-13C_A/(6\beta_0)} \,\Big]
 + a_5 \, \pi\alpha_s(m_t) \Big[\,1- z^{1-2 C_A/\beta_0} \, \Big] 
\nn \\[2mm]  && 
{}+ a_0\, \pi\alpha_s(m_t)) \Big[\, (z-1)-w^{-1}\ln(w) \,\Big] \,,
\end{eqnarray}
with
\begin{equation*}
z = \frac{\alpha_s(m_t\,\nu)}{\alpha_s(m_t)},\qquad\qquad 
w = \frac{\alpha_s(m_t\,\nu^2)}{\alpha_s(m_t\,\nu)},
\end{equation*}
and
\begin{eqnarray}
  a_2 &=& 
   \frac{ C_F [ C_A\,C_F ( 9 C_A - 100 C_F) 
   - \beta_0\,( 26 C_A^2 + 19 C_A C_F - 32 C_F^2 ) ]}{ 26\,\beta_0^2\, C_A }
   \,, \nn\\[2mm]
  a_3 &=&
    \frac{C_F^2 }{ \beta_0^2\, (6\beta_0-13C_A) (\beta_0-2C_A)}
    \, \Big\{ C_A^2 ( 9 C_A- 100 C_F )
    + \beta_0\,C_A \Big[ 74 C_F + C_A ( 13 {\bf S^2} - 42 ) \Big]\nn \\[2mm]
   &&\qquad - 6 \beta_0^2  \Big[ 2 C_F + C_A ( {\bf S^2} - 3 ) \Big]\,\Big\} 
   \,,\nn\\[2mm]
  a_4 &=& \frac{24 C_F^2 (11 C_A\!-\!3\beta_0)(5 C_A\!+\!8 C_F) }{ 13\, C_A
     (6\beta_0\!-\!13 C_A)^2},\quad
  a_5 = \frac{C_F^2 \big[ C_A( 15\! -\! 14\,{\bf S^2} ) 
    \!+\! \beta_0 (4{\bf S^2}\!-\!3) \big] }{ 6 (\beta_0\!-\!2 C_A)^2 }\,,
\\[2mm]
  a_0 & = & -\frac{8\,C_F\,(C_A+C_F)\,(C_A+2\,C_F)}{3\beta_0^2} \,.
\end{eqnarray}
The term ${\bf S^2}$ is the square of the total $t\bar t$ spin. We take the
convention where the matching conditions at $\nu=1$ only account for QCD
effects, so at LL order we have $c_{1}(1)=c_{0}(1)=1$. 

Through the optical theorem the $t\bar t$ production rate for a $t\bar t$
invariant mass $Q^2\approx 4m_t^2$ involves the imaginary part of the
time-ordered product of the production and annihilation currents defined in
Eqs.~(\ref{J1J0}),
\begin{eqnarray}
 {\cal A}_1^{lk}(Q^2,m_t,\nu) & = & i\,
 \sum\limits_{\mbox{\scriptsize\boldmath $p$},\mbox{\scriptsize\boldmath $p'$}}
 \int\! d^4x\: e^{-i \hat{q} \cdot x}\:
 \Big\langle\,0\,\Big|\, 
    T\, J^{l\dagger}_{1,{\sitbf{p}^\prime}}(0) J^k_{1,{\sitbf{p}}}(x)
 \Big|\,0\,\Big\rangle 
\,=\,2 N_c \delta^{lk}\, G^c(a,v,m_t,\nu)\,,\quad
\label{A1def}
\\[2mm]
 {\cal A}_0(Q^2,m_t,\nu) & = & i\,
 \sum\limits_{\mbox{\scriptsize\boldmath $p$},\mbox{\scriptsize\boldmath $p'$}}
 \int\! d^4x\: e^{-i \hat{q} \cdot x}\:
 \Big\langle\,0\,\Big|\, 
    T\, J^\dagger_{0,{\sitbf{p}^\prime}}(0) J_{0,{\sitbf{p}}}(x)
 \Big|\,0\,\Big\rangle 
\, = \, N_c\, G^c(a,v,m_t,\nu)\,,
\label{A2def}
\end{eqnarray}
where $\hat{q}\equiv(\sqrt{Q^2}-2m_t,0)$ and 
\begin{eqnarray}
v & = &
\sqrt{\frac{\sqrt{Q^2}-2 m_t-2\delta m_t+i\Gamma_t}{m_t}}
\,,
\label{vdef}
\end{eqnarray}
is the c.m.\,\,top quark relative velocity. The term $G^c$ is the zero-distance S-wave
Coulomb Green function of the nonrelativistic Schr\"odinger equation with the 
potential in Eq.~(\ref{VCoulomb}). At LL order (i.e. including only the first term on the RHS of Eq.~\eqref{VCoulomb}~) the Green function has a simple
analytic form and reads (in dimensional regularization)~\cite{hmst}
\begin{eqnarray}
 G^c_{\rm LL}(a,v,m_t,\nu) & = &
 \frac{m_t^2}{4\pi}\left\{\,
 i\,v - a\left[\,\ln\left(\frac{-i\,v}{\nu}\right)
 -\frac{1}{2}+\ln 2+\gamma_E+\psi\left(1\!-\!\frac{i\,a}{2\,v}\right)\,\right]
 \,\right\}
 \nonumber \\[2mm] & &
 + \,\frac{m_t^2\,a}{4 \pi}\,\,\frac{1}{4\,\epsilon}
\,.
\label{deltaGCoul}
\end{eqnarray}  
Explicit analytic expressions exist for the NLL order correction to $G^c$
coming from one insertion of the ${\cal O}(\alpha_s^2)$ contributions to the
effective Coulomb potential in Eq.~(\ref{VCoulomb})~\cite{Penin1}. We
use the numerical techniques and codes of the TOPPIC program developed in
Ref.~\cite{Jezabek1} (see also Ref.~\cite{Strassler1}) and  
determine an exact solution of the full NLL Schr\"odinger equation following
the approach of Refs.~\cite{hmst}.  

\section{Higgs Energy Distribution at LL Order} 
\label{sectionLL}

\begin{figure}[t] 
  \begin{center}
    \epsfig{file=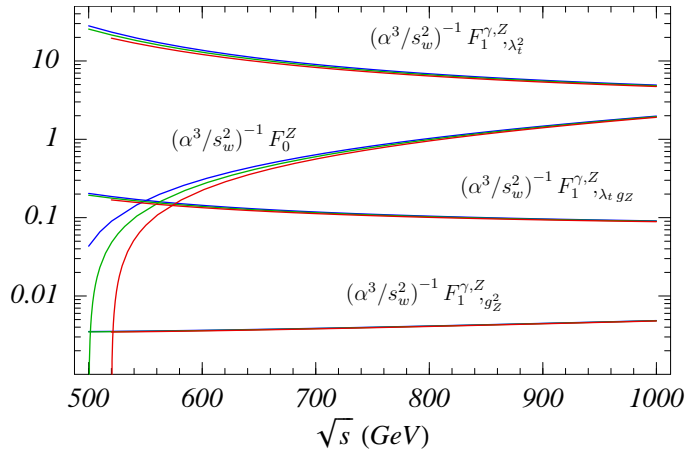,height=6cm}
    \caption{The singlet form factor $F_0^{Z}$ and the different
    contributions to the triplet form factor $F_1^{\gamma,Z}$ for
    $m_H=120$ GeV (respective upper lines), $m_H=140$ GeV (respective 
    middles lines), and $m_H=160$ GeV (respective lower lines). Here, 
    $F^{\gamma,Z}_{1,\lambda_t^2}$ refers to the
    contribution to $F_1^{\gamma,Z}$ which is proportional to
    $\lambda_t^2$, and 
    $F^{\gamma,Z}_{1,\lambda_t\,g_Z}$ and $F^{\gamma,Z}_{1,g_Z^2}$
    are defined analogously. The top mass is set to $m_t=180$~GeV and the
    other parameters are given in Eqs.~\eqref{parameters}. 
      \label{fig2} }
  \end{center}
\end{figure}

To illustrate the method of computing the Higgs energy distribution for large
Higgs energies we start by considering the simplified case where the process is
mediated through a virtual photon only. The amplitude for the process 
\begin{equation}
e^+(k^\prime)\, e^-(k)  \to \, \gamma^* \, \to \,
t(k_1)\, \bar t(k_2)\, H(p_H)
\end{equation}
in the full theory reads
\begin{eqnarray}
{\cal M}_\gamma & = & i\,\lambda_t\,\frac{4\pi Q_e Q_t\alpha}{s}\,
\left[\, \bar v_{e^+}(k) \gamma_\mu u_{e^-}(k^\prime) \, \right]
\nonumber\\ &&
\times\,\left[\,
\bar u_t(k_1)\left(
\frac{\slash{p}_H+\slash{k_1}+m_t}{(p_H+k_1)^2-m_t^2}\gamma^\mu +
\gamma^\mu \frac{-\slash{p}_H-\slash{k_2}+m_t}{(p_H+k_2)^2-m_t^2}
\right)\,v_{\bar t}(k_2)
\,\right]\,,
\end{eqnarray}
where
\begin{eqnarray}
\lambda_t & = & \frac{e}{2 s_w}\frac{m_t}{M_W}
\,,
\end{eqnarray}
$e$ being the electric charge and $s_w$ ($c_w$) the sine (cosine) of the Weinberg angle.
In the large Higgs energy region one has 
\begin{eqnarray}
E_H & \approx & E_H^{\rm max} \, = \, \frac{1}{2\sqrt{s}}
\left(s+m_H^2-4m_t^2\right)
\nonumber \\[2mm]
k_1 & \approx & k_2 \, \approx \, k_t \,=  \, (E_t,\bmk_t) \,,
\nonumber \\[2mm]
E_t & \equiv & \sqrt{m_t^2+|\bmk_t|^2}
\,,\qquad\quad
|\bmk_t| \, \equiv \, \frac{1}{2}\, 
\left( \frac{(s+m_H^2-4m_t^2)^2}{4s} -m_H^2 \right)^{1/2} 
\,,
\end{eqnarray}
where
\begin{eqnarray}
s & = & (k+k^\prime)^2
\,,
\end{eqnarray}
and one can relate the effective theory top and antitop spinors in the $t\bar
t$ c.m.\,\,frame with the full theory spinors by a Lorentz boost,
\begin{eqnarray}
u_{t}(k_1) & = & U(\bmk_t)\,\left(
\begin{array}{c}\psi_{{\sitbf{p}}}\\0\end{array}\right)
\,=\, \left(\frac{E_t+m_t}{2m_t}\right)^{1/2}\,
\left(
\begin{array}{c}\psi_{{\sitbf{p}}}\\
\frac{\sbsigma\bmk_t}{E_t+m_t}\psi_{{\sitbf{p}}}\end{array}\right)
\,,
\nonumber\\[2mm]
v_{\bar t}(k_2) & = & i\gamma^2\,
\left[\,U(\bmk_t)\,\left(
\begin{array}{c}\chi_{-{\sitbf{p}}}\\0\end{array}\right)
\,\right]^*
\,=\,  
\left(\frac{E_t+m_t}{2m_t}\right)^{1/2}\,
\left(
\begin{array}{c}i \sigma_2 \chi^*_{{\sitbf{p}}}\\
\frac{\sbsigma\bmk_t}{E_t+m_t}(i \sigma_2\chi^*_{{\sitbf{p}}})
\end{array}\right)
\,,
\label{matchspinors}
\end{eqnarray}
where $\itbf{p}$ is the small residual top quark momentum in the $t\bar t$
c.m.\,\,frame. Note that on the RHS's shown 
in Eqs.~(\ref{matchspinors}) we have only kept the effective theory operators 
that are leading order in the expansion for $\itbf{p}\ll m_t$ in the $t\bar t$
rest frame. This is sufficient for the matching computation at LL and NLL
order. We also note that in the effective theory we describe the
antitop quark by a positive energy spinor which makes the charge conjugation
operation in the second equation necessary. In the large Higgs energy end point
region the amplitude then reduces to the form 
\begin{eqnarray}
{\cal M}_\gamma(E_H\approx E_H^{\rm max}) & = &  
i\,\lambda_t\,\frac{4\pi Q_e Q_t\alpha}{s}\,
\Big[\,\bar v_{e^+}(k^\prime)\,\gamma_i \, u_{e^-}(k)\,\Big]\,
\Big[\,\psi_{{\sitbf{p}}}^\dagger\, \sigma_j (i\sigma_2)
  \chi_{-{\sitbf{p}}}^*\,\Big]
\, T^{ij}
\,,
\nonumber\\[2mm]
 T^{ij} & = &
\frac{2\sqrt{s}}{(p_H+k_t)^2-m_t^2}\,
\left[\, \frac{E_t}{m_t}\delta_{ij} - \frac{k_t^i k_t^j}{m_t(E_t+m_t)}
\,\right]
\,.
\end{eqnarray}
We see that for virtual photon exchange the top-antitop pair is produced only
in the spin triplet configuration.
Without QCD corrections, the distribution in the large Higgs energy
region then reads 
\begin{eqnarray}
d\sigma & = &
\frac{(2m_t)^2}{2s(2\pi)^5}\,\frac{1}{4}\sum_{\rm{spins}}\,
\left|{\cal M}_\gamma(E_H\approx E_H^{\rm max})\right|^2\,
\delta^{(4)}(k+k^\prime-k_1-k_2-p_H)\,
\nonumber\\[2mm] & &\hspace{3cm} \times\,
\delta_+(p_H^2-m_H^2)\,
\delta_+(k_1^2-m_t^2)\,
\delta_+(k_2^2-m_t^2)\,d^4p_H\,d^4k_1\,d^4k_2\,
\nonumber\\[4mm] & = &
\frac{\lambda_t^2\, Q_e^2\, Q_t^2\, \alpha^2}{8\pi s^{5/2}}\,
L_{ij}\,T_{ik}\,T^*_{jl}\,
\mbox{Im}\left[ \tilde{\cal A}_1^{lk}((k+k^\prime-p_H)^2,m_t)\right]\,
\nonumber\\[2mm] & & \hspace{3cm}\times\,
\left[(1+x_H-4x_t)^2-4x_H\right]^{1/2}\,d E_H\,d\Omega_H
\,,
\nonumber\\[4mm] 
L_{ij} & = & 2 s \left(\delta^{ij} - \frac{k^ik^j}{\bmk^2}\right)
\,,
\end{eqnarray}
where $\delta_+$ indicates that in the on-shell condition only
positive energies are accounted for. 
The term $\tilde{\cal A}_1^{lk}$ is the current correlator in Eq.~(\ref{A1def}) 
without QCD effects and for stable top quarks, $L_{ij}$ the lepton tensor and 
\begin{eqnarray}
x_t & \equiv & \frac{m_t^2}{s}\,,\qquad 
x_H \, \equiv \, \frac{m_H^2}{s}\,,\qquad
x_Z \, \equiv \, \frac{m_Z^2}{s}
\,.
\end{eqnarray}
In the large Higgs energy region,
including the QCD effects coming from the Coulomb potential in
Eq.~(\ref{VCoulomb}) and the finite top quark lifetime, the LL result
for the Higgs energy distribution reads  
\begin{eqnarray}
d\sigma & = & \frac{8\,N_c\,F^\gamma}{s^{3/2}\,m_t^2}
\mbox{Im}\left[\, G_{\rm LL}^c(a,v,m_t,\nu)\,\right]\,
\left[(1+x_H-4x_t)^2-4x_H\right]^{1/2}\,d E_H
\,,
\nonumber\\[4mm] 
F^\gamma & = &
\frac{\lambda_t^2\, Q_e^2 \,Q_t^2\,\alpha^2}{3}\,\frac{(1-x_H+4x_t)^2+8x_t}{(1+x_H-4x_t)^2}
\,.
\end{eqnarray}
If the virtual Z-exchange is accounted for, the computation is slightly
more involved since we also have to account for the top pair production in the
singlet spin state configuration and both correlators in
Eqs.~(\ref{A1def},\ref{A2def}) are needed. Including the Wilson coefficient of the
currents in Eqs.~(\ref{currentWilson}) as well, the result can be written in the form
\begin{eqnarray}
\frac{d\sigma}{d E_H}(E_H\approx E_H^{\rm max}) & = &
\frac{8\,N_c\,\left[(1+x_H-4x_t)^2-4x_H\right]^{1/2}}{s^{3/2}\,m_t^2}\,
\left(\,c^2_0(\nu)\, F^Z_0 + c_1^2(\nu)\,F^{\gamma,Z}_1\,\right)\,
\nonumber\\[2mm] & & \hspace{1cm}
\times\,\mbox{Im}\left[\, G^c(a,v,m_t,\nu)\,\right]\,
\,,
\label{dsdEHEFT}
\end{eqnarray}
where
\begin{eqnarray}
F_1^{\gamma,Z} & = & 
\frac{\alpha^2 \lambda_t^2}{3}\,
\frac{(1 - x_H + 4x_t)^2 + 8x_t}{(1 + x_H - 4x_t)^2}\,
 \left(\, Q_e^2 Q_t^2  + \frac{v_t^2\left(a_e^2 + v_e^2\right)}{(1 - x_Z)^2} 
  + \frac{2 Q_e Q_t v_e v_t}{1 - x_Z} \right)
\nonumber\\[2mm] & &
+ \, 2 \alpha^2 g_Z \lambda_t\, 
\frac{(x_t x_Z)^{1/2}(1 - x_H + 4x_t)}{(1 + x_H - 4x_t)(4x_t - x_Z)(1 - x_Z)}
\,\left(  \frac{v_t^2 \left(a_e^2 + v_e^2\right)}{(1 - x_Z)} + Q_e Q_t v_e v_t
\right) 
\nonumber\\[2mm] & &
+ \, \frac{\alpha^2 g_Z^2 v_t^2\left(a_e^2 + v_e^2\right)}{24}\, 
  \frac{((1 - x_H + 4x_t)^2 + 32x_t)x_Z}{(4x_t - x_Z)^2(1 - x_Z)^2}
\,,
\\[4mm]
F_0^{Z} & = &
\frac{\alpha^2 g_Z^2 a_t^2\left(a_e^2 + v_e^2\right)}{24}\,  
\frac{(1 - x_H + 4x_t)^2 - 16 x_t}{(1 - x_Z)^2\,x_Z}
\,,
\end{eqnarray}
and
\begin{eqnarray}
  v_f = \frac{T_3^f-2 Q_f s_w^2}{2s_w c_w}\,,
  \qquad
  a_f = \frac{T_3^f}{2s_w c_w} \,,
  \qquad
  g_Z = \frac{e}{2 s_w c_w} \,.
\end{eqnarray}
Here $Q_f$ is the charge for fermion $f$, and $T_3^f$ is the third
component of weak isospin.  
In Fig.~\ref{fig2} it is shown that the singlet
contribution governed by the coefficient $F_0^{Z}$ is negligible for
c.m.\,energies around $500$~GeV. For larger c.m.\,energies it becomes
more important, but remains smaller than the triplet contribution for
all available energies. Figure~\ref{fig2} also illustrates that the
triplet contribution, determined by the coefficient $F_1^{\gamma,Z}$,
is dominated by the term proportional to the square of the Yukawa
coupling $\lambda_t$. This term describes the effects of Higgs
radiation off one of the top quarks. 

%
%

We note that the NLL (${\cal O}(\alpha_s)$)
matching conditions for the three triplet Wilson coefficients
$c_{1,j}$ depend on 
the $t\bar t$ spin configuration (i.e.\,\,on $j$) since the kinematic
situation for $t\bar t H$ production in the large Higgs energy end point is not
invariant under separate rotations of the spin quantization axis. However, for
our purposes it is sufficient to define a triplet Wilson coefficient that is
averaged over the three spin configurations. Using such an averaged triplet
Wilson coefficient the Higgs energy spectrum at NLL order can also be cast in
the simple form of Eq.~(\ref{dsdEHEFT}). 

\section{Matching Conditions at NLL Order} 
\label{sectionmatching}

At NLL order, we need
to account for the  ${\cal O}(\alpha_s^2)$ contributions to the Coulomb
potential in Eq.~(\ref{VCoulomb}), the NLL running of the coefficients $c_1$
and $c_0$ and their ${\cal O}(\alpha_s)$ matching conditions at $\nu=1$. The
latter QCD corrections are process specific and cannot be inferred from
results obtained in earlier computations for the $t\bar t$ threshold in
$e^+e^-\to t\bar t$.    
%
%
\begin{figure}[t] 
\begin{center}
 \leavevmode
 \epsfxsize=14cm
 \leavevmode
 \epsffile[130 300 460 450]{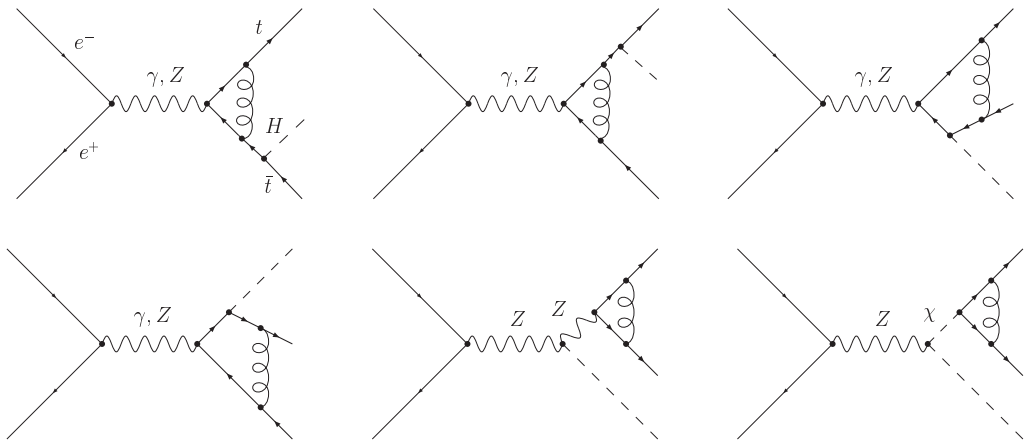}
 \vskip  0.0cm
 \caption{
Diagrams describing virtual one-loop QCD corrections in the Standard Model to
compute the NLL matching conditions for the operators  $J_{1,{\sitbf{p}}}$ and
$J_{0,{\sitbf{p}}}$ that describe $t\bar t$ production for invariant masses
$Q^2\approx 4m_t^2$. Self energy diagrams are implied.
 \label{fig3} }
\end{center}
\end{figure}
Since we are interested in the region where $Q^2\approx 4m_t^2$ and we can
neglect gluon exchange with the Higgs final state, we can apply the same
nonrelativistic powercounting as the one known for the process $e^+e^-\to t\bar
t$. Thus real gluon radiation, which is related to emission of
gluons with energies of order $m_t v^2\sim m_t\alpha_s^2$ in the $t\bar t$
c.m.\,frame, is suppressed by a factor
$\alpha_s\times(Q^2-4m_t^2)/m^2\sim v^3$ (using the nonrelativistic
counting $v\sim\alpha_s$). Therefore 
we only need to take into account the hard contributions contained in the
virtual QCD corrections shown in Fig.~\ref{fig3}. We have extracted the hard QCD
corrections from the results for the Standard Model amplitude for 
$e^+e^-\to t\bar t H$ computed in Ref.~\cite{Denner1}. The results of 
Ref.~\cite{Denner1} were given in term of numerical codes for form
factors to 
a number of standard matrix elements. We have computed the contributions of
these form factors to $d \sigma/d E_H$ in the large Higgs energy region. To
the result we have matched the corresponding expression for the ${\cal
  O}(\alpha_s)$ corrections contained in the effective theory expression 
in Eq.~(\ref{dsdEHEFT}) at $\mu=m_t$ ($\nu=1$). This allowed us to determine
the ${\cal O}(\alpha_s)$ matching conditions to $c_{0,1}$ numerically,
\begin{equation}
c_{0,1}(\nu=1) = 1 + \frac{C_F\alpha_s(m_t)}{2}\,\delta
c_{0,1}(\sqrt{s},m_t,m_H)
\,.
\label{matchcond}
\end{equation}
\begin{table}
\begin{center}
\begin{tabular}{|c|c||c|c|c|c|c|c|}
\hline
\multicolumn{2}{|c|}{$\delta c_1$} &
\multicolumn{6}{|c|}{$\sqrt{s}$}\\\hline\hline
$\quad m_t\quad$ & $\quad m_H\quad$ & $500$ & $600$ & $700$ & $800$ & $900$ & $1000$ \\\hline\hline 
$175$ & $120$ & $-2.266(1)$ & $-2.339(1)$ & $-2.397(1)$ & $-2.453(1)$ &
$-2.509(1) $ & $-2.564(1)$  \\ \hline
 & $140$ & & $-2.339(1)$ & $-2.401(1)$ & $-2.459(1)$ & $-2.516(1)$ &
$-2.572(1)$  \\ \hline
 & $160$ & & $-2.340(0)$ & $-2.407(1)$ & $-2.467(1)$ & $-2.524(1)$ &
$-2.581(1)$ \\ \hline
$180$ & $120$ & $-2.254(1)$ & $-2.329(1)$ & $-2.386(1)$ & $-2.441(0)$ &
$-2.495(1) $ & $-2.548(1)$  \\ \hline
 & $140$ & & $-2.327(1)$ & $-2.389(1)$ & $-2.446(1)$ & $-2.501(1)$ &
$-2.555(1)$  \\ \hline
 & $160$ & & $-2.328(1)$ & $-2.394(1)$ & $-2.452(0)$ & $-2.509(0)$ &
$-2.564(1)$  \\ \hline
\hline
\end{tabular}
\end{center}
{\tighten \caption{Numerical values for the averaged spin triplet matching
condition $\delta c_1(\sqrt{s},m_t,m_H)$ for typical values of
$\sqrt{s}$, $m_t$ and $m_H$. The masses and energies are given in units of GeV.}
\label{tab1} }
\end{table}
\begin{table}
\begin{center}
\begin{tabular}{|c|c||c|c|c|c|c|c|}
\hline
\multicolumn{2}{|c|}{$\delta c_0$} &
\multicolumn{6}{|c|}{$\sqrt{s}$}\\\hline\hline
$\quad m_t\quad$ & $\quad m_H\quad$ & $500$ & $600$ & $700$ & $800$ & $900$ & $1000$ \\\hline\hline 
$175$ & $120$ & $-0.562(5)$ & $-0.638(4)$ & $-0.692(4)$ & $-0.733(1)$ &
$-0.763(2) $ & $-0.787(1)$  \\ \hline
 & $140$ & & $-0.636(3)$ & $-0.691(4)$ & $-0.732(3)$ & $-0.762(2)$ &
$-0.786(1)$  \\ \hline
 & $160$ & & $-0.634(3)$ & $-0.690(4)$ & $-0.731(4)$ & $-0.762(1)$ &
$-0.785(2)$ \\ \hline
$180$ & $120$ & $-0.553(4)$ & $-0.627(3)$ & $-0.684(2)$ & $-0.725(1)$ &
$-0.756(1) $ & $-0.780(2)$  \\ \hline
 & $140$ & & $-0.625(4)$ & $-0.682(4)$ & $-0.725(4)$ & $-0.755(2)$ &
$-0.780(2)$  \\ \hline
 & $160$ & & $-0.622(4)$ & $-0.681(4)$ & $-0.723(1)$ & $-0.755(1)$ &
$-0.779(2)$  \\ \hline
\hline
\end{tabular}
\end{center}
{\tighten \caption{Numerical values for the spin singlet matching
condition $\delta c_0(\sqrt{s},m_t,m_H)$ for typical values of
$\sqrt{s}$, $m_t$ and $m_H$. The masses and energies are given in units of GeV.}
\label{tab2} }
\end{table}
With this method the averaged triplet coefficient is obtained automatically.
Since the numerical results of Ref.~\cite{Denner1} are functions of the
four-momenta of the initial and final state particles we were able to approach
the large Higgs energy end point from several directions in the final state
phase space in order to test the numerical stability and the
independence of the result on the direction with which the end point
is approached.  For example, depending on the top-antitop quark
c.m.\,\,relative velocity $v$, the Higgs and the top and antitop quark four
momenta (moving in the x-y plane to be definite) can be written as 
\begin{eqnarray}
p_H & = & (E_H,0,k_H,0) \,,
\nonumber\\[2mm]
k_1 & = & (E_1,-|\bmk_1|\sin\theta,|\bmk_1|\cos\theta,0)\,,
\nonumber\\[2mm]
k_2 & = & (E_2,-|\bmk_2|\sin(\bar\theta+\theta),
|\bmk_2|\cos(\bar\theta+\theta),0)\,,
\end{eqnarray} 
where 
\begin{eqnarray}
|\bmk_{i}| & = & \sqrt{E_{i}-m_i^2}\,,\qquad
E_H \, = \, \frac{1}{2\sqrt{s}}\left(s+m_H^2-Q^2\right) \,,\qquad
Q^2 \,=\, \frac{4m_t^2}{1-v^2}\,,
\nonumber\\[2mm]
E_1 & = & \sqrt{s}-E_H-E_2\,,\qquad
E_2 \,=\, \frac{1}{2}\left(\sqrt{s}-E_H-\frac{\sqrt{s}}{2}\beta\right)\,, 
\nonumber\\[2mm]
\beta & = & \frac{x}{\sqrt{s}}\left(
\frac{4(E_H^2-m_H^2)\,(s-2\,\sqrt{s}\,E_H+m_H^2-4m_t^2)}{(s-2\,\sqrt{s}\,E_H+m_H^2)}
\right)^{1/2}\,, 
\nonumber\\[2mm]
\bar\theta & = & \arccos\left[\,
\frac{s + 2m_t^2 - 2\,\sqrt{s}\,(E_1 + E_2) + 2 \,E_1\, E_2 - m_H^2}
{2|\bmk_1||\bmk_2|}
\,\right]
\,, 
\nonumber\\[2mm]
\theta & = &
\arccos\left[\, 
\frac{E_H\, E_1 - \sqrt{s}\, E_1 + m_t^2 + E_1\, E_2 - |\bmk_1||\bmk_2|\,\cos\bar\theta}
{|\bmk_H||\bmk_1|} 
\,\right]
\,.
\end{eqnarray}
The variable $x$ can be chosen between $0$ and $1$ and interpolates between
the two extreme situations $x=0$, which refers to the symmetric configuration
where $E_1=E_2$, and $x=1$, which describes the situation where the top and
antitop quarks move into the same direction having the maximal possible energy
difference. We have extracted the functions $\delta c_{0,1}$ numerically in
the limit $v\to 0$ and estimated the uncertainties from extrapolating to $v=0$
and from varying the variable $x$ between $0.1$ and $0.9$. For $x$ too close 
to zero or one the extraction procedure becomes unreliable due to
numerical instabilities that develop in the code of
Ref.~\cite{Denner1} in these particular edges of the phase space. 
In order to identify the singlet and the triplet contributions in the full
theory we used the helicity basis for the top and antitop spinors in the large 
Higgs energy end point where $k_1=k_2$ to construct the singlet and the three
triplet $t\bar t$ spin configurations from the standard matrix elements given in
Ref.~\cite{Denner1}.  The results (including our
error estimate) for some
representative values for $\sqrt{s}$, $m_t$ and $m_H$ are shown in
Tab.~\ref{tab1} for $\delta c_1$ and in Tab.~\ref{tab2} for $\delta c_0$.
In particular, for smaller c.m.\,energies we find that the absolute
uncertainties for the singlet matching conditions are a few times larger than
for the triplet matching conditions. This is a numerical effect that can be
understood from the fact that for smaller c.m.\,energies the contributions
coming from the singlet $t\bar t$ spin configuration are much smaller than
those coming from the triplet spin configuration.  In both cases, however, the
relative uncertainties are well below 1\%.

\section{Numerical Analysis} 
\label{sectionanalysis}
%
%

%
%
\begin{figure}[t] 
\begin{center}
 \leavevmode
 \epsfxsize=10cm
 \leavevmode
 \epsffile[100 430 580 730]{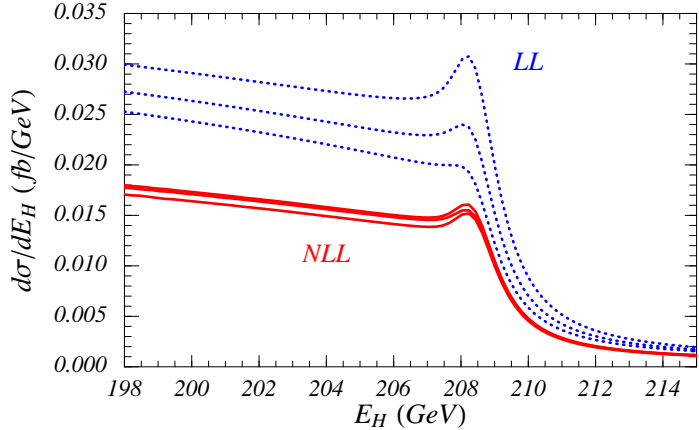}
 \vskip  -0.7cm
 \caption{
Higgs energy spectrum in the end point region at LL (dotted lines) and 
NLL (solid lines) order in the nonrelativistic expansion for
the vNRQCD renormalization parameters $\nu=0.1,0.2,0.4$ and for
$\sqrt{s}=600$~GeV, $m_t^{\rm 1S}=180$~GeV, $m_H=140$~GeV.
 \label{fig4} }
\end{center}
\end{figure}
%
%

%
%
\begin{figure}[t] 
\begin{center}
 \leavevmode
 \epsfxsize=8cm
 \leavevmode
 \epsffile[100 430 540 730]{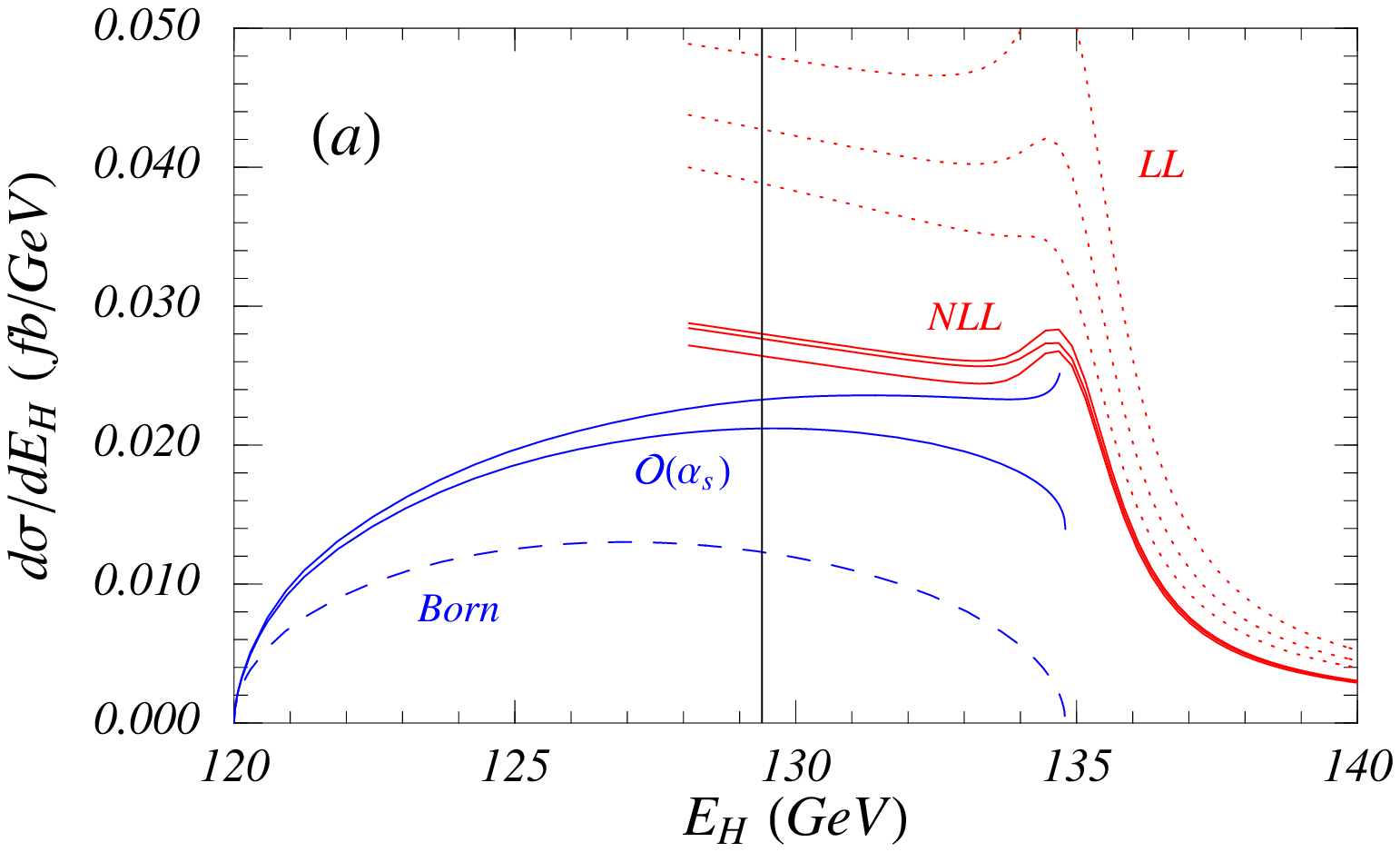}
 \leavevmode
 \epsfxsize=8cm
 \leavevmode
 \epsffile[100 430 540 730]{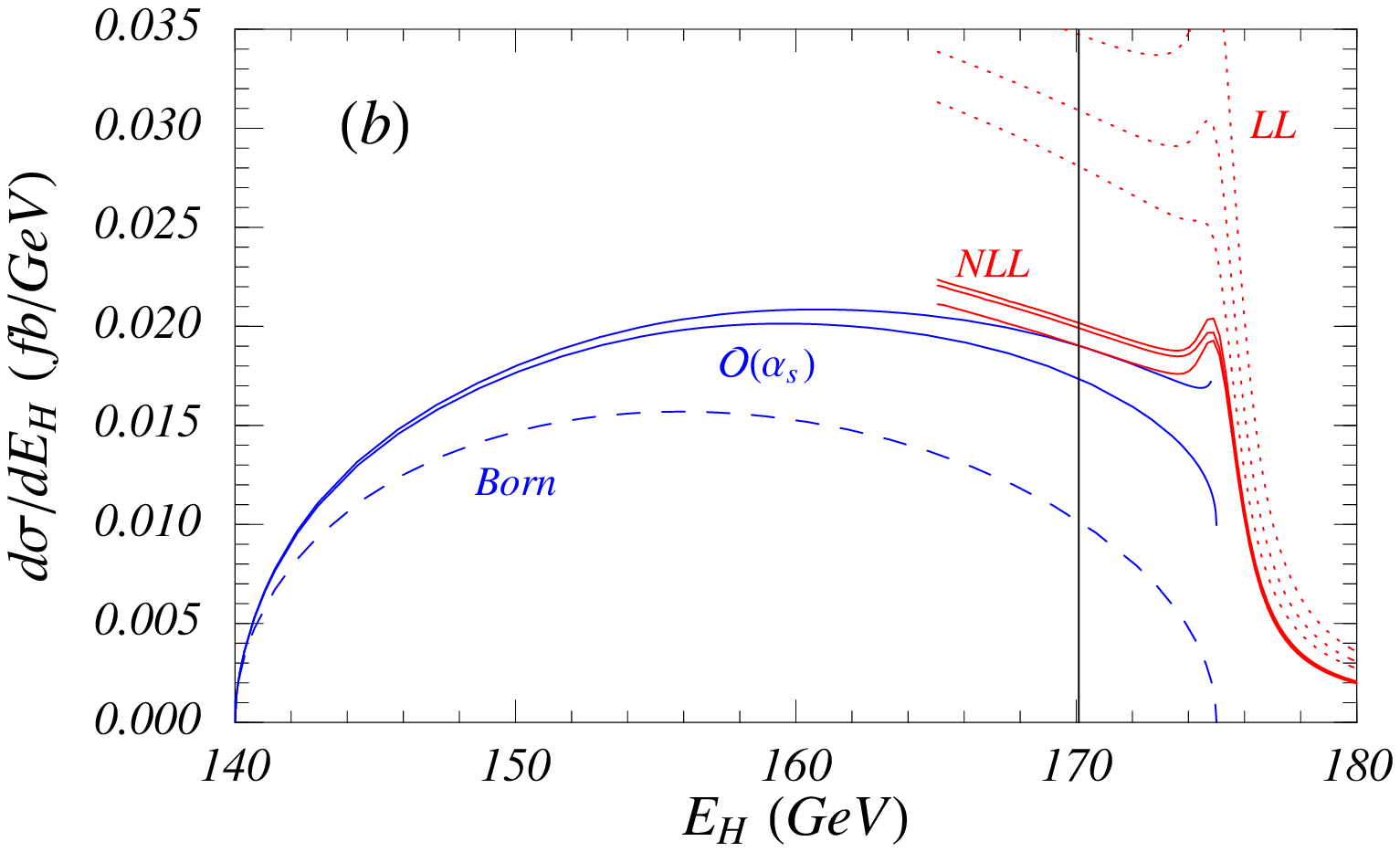}\\[-5mm]
 \leavevmode
 \epsfxsize=8cm
 \leavevmode
 \epsffile[100 430 540 730]{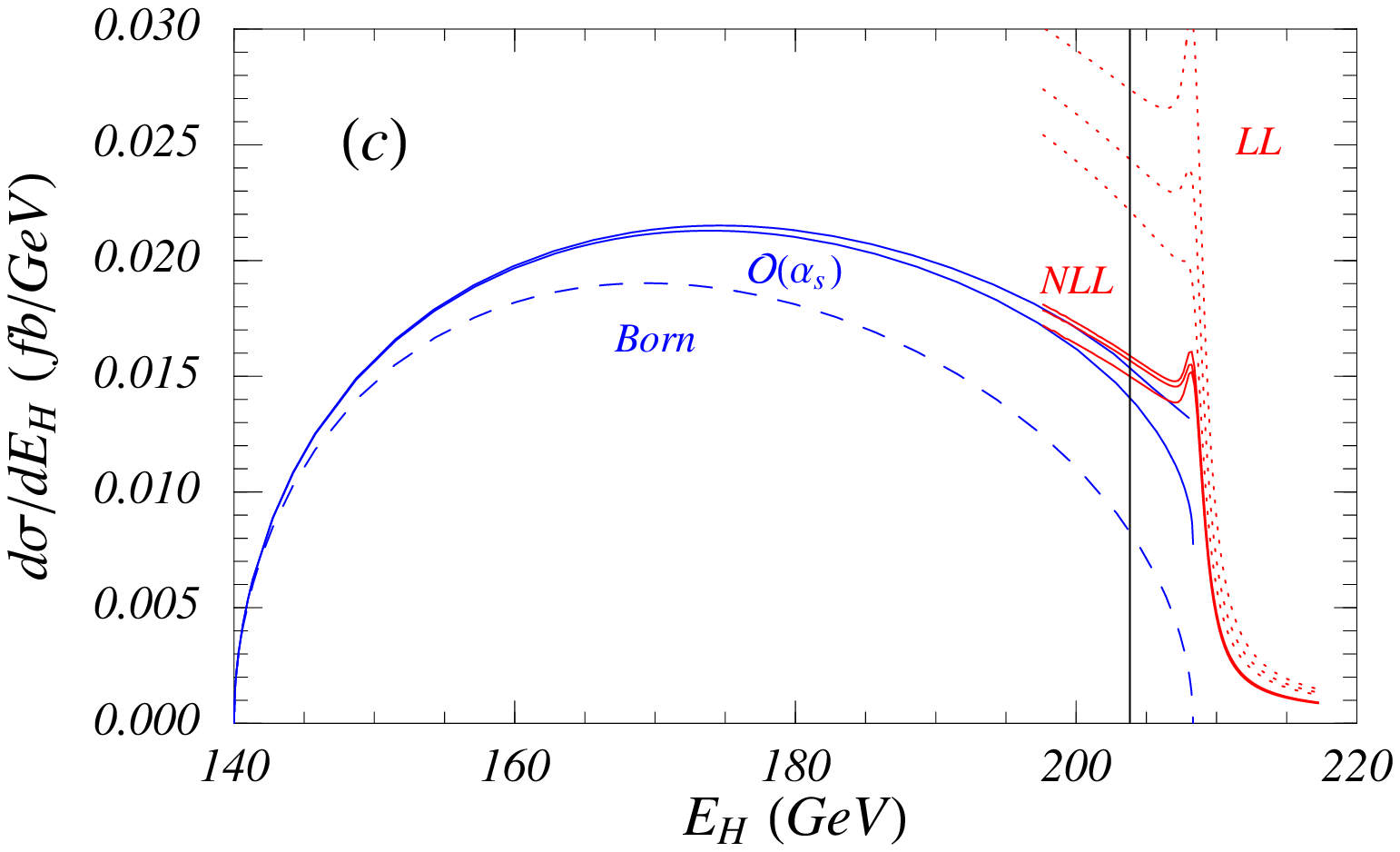}
 \leavevmode
 \epsfxsize=8cm
 \leavevmode
 \epsffile[100 430 540 730]{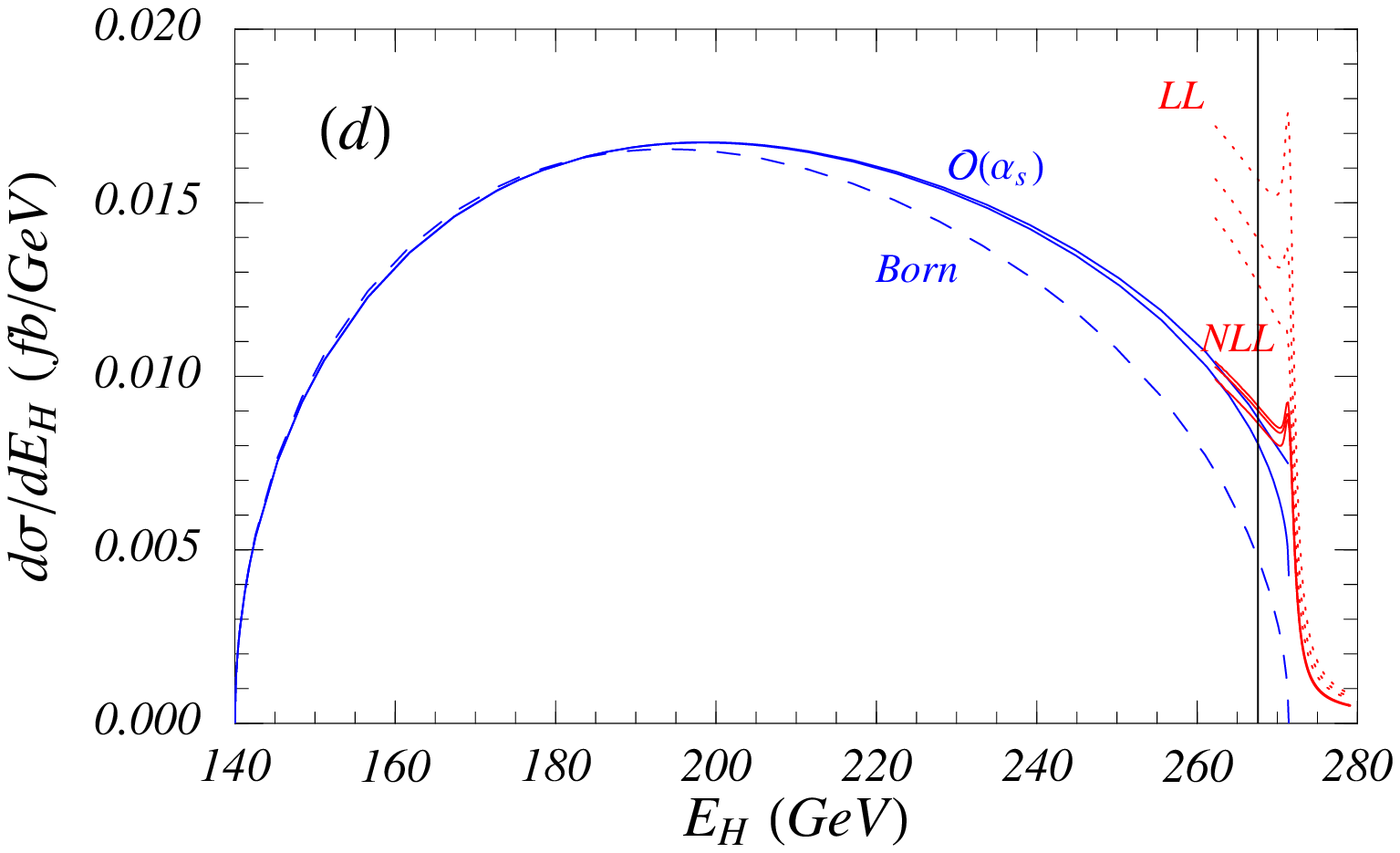}
 \vskip  -0.8cm
 \caption{
Higgs energy spectrum at LL (dotted lines) and 
NLL (solid lines) order in the nonrelativistic expansion
for the vNRQCD renormalization parameters $\nu=0.1,0.2,0.4$
and at the Born level and at ${\cal O}(\alpha_s)$ for
$\mu=\sqrt{s}, \sqrt s |v|$ for the parameters
(a) $\sqrt{s}=500$~GeV, $m_H=120$~GeV,
(b) $\sqrt{s}=550$~GeV, $m_H=140$~GeV,
(c) $\sqrt{s}=600$~GeV, $m_H=140$~GeV,
(d) $\sqrt{s}=700$~GeV, $m_H=140$~GeV.
For all cases that top mass has been set to
$m_t^{\rm 1S}=180$~GeV
 \label{fig5} }
\end{center}
\end{figure}
\begin{table}
\begin{center}
\begin{tabular}{|c|c||c|c||l||c|c|c|}
\hline
%
$\sqrt{s}$ [GeV] & 
$m_H$ [GeV] & 
$\mbox{}\;\sigma(\mbox{Born})$ [fb] & 
$\mbox{}\;\sigma(\alpha_s)$ [fb] &
$\mbox{}\;\sigma(\mbox{NLL})$ [fb] & 
$\mbox{}\quad\frac{\sigma(\mbox{\tiny NLL})}{\sigma(\mbox{\tiny Born})}\quad\mbox{}$ &
$\mbox{}\quad\frac{\sigma(\mbox{\tiny NLL})}{\sigma(\alpha_s)}\quad\mbox{}$ &
$\frac{\sigma(\mbox{\tiny NLL})_{|\beta|<0.2}}{\sigma(\alpha_s)_{\beta<0.2}}$
\\ \hline\hline
$500$ & $120$ & $ 0.151$ & $ 0.263$ & $\quad 0.357(20)$ & $ 2.362$ & $ 1.359$ & $ 1.78$\\\hline
$550$ & $120$ & $ 0.984$ & $ 1.251$ & $\quad 1.342(37)$ & $ 1.364$ & $ 1.073$ & $ 1.66$\\\hline
$550$ & $140$ & $ 0.430$ & $ 0.589$ & $\quad 0.658(24)$ & $ 1.530$ & $ 1.117$ & $ 1.68$\\\hline
$550$ & $160$ & $ 0.134$ & $ 0.207$ & $\quad 0.254(12)$ & $ 1.902$ & $ 1.226$ & $ 1.74$\\\hline
$600$ & $120$ & $ 1.691$ & $ 1.939$ & $\quad 2.005(30)$ & $ 1.185$ & $ 1.034$ & $ 1.66$\\\hline
$600$ & $140$ & $ 1.014$ & $ 1.203$ & $\quad 1.259(24)$ & $ 1.241$ & $ 1.046$ & $ 1.67$\\\hline
$600$ & $160$ & $ 0.565$ & $ 0.700$ & $\quad 0.745(18)$ & $ 1.319$ & $ 1.065$ & $ 1.68$\\\hline
$700$ & $120$ & $ 2.348$ & $ 2.454$ & $\quad 2.485(13)$ & $ 1.058$ & $ 1.012$ & $ 1.68$\\\hline
$700$ & $140$ & $ 1.695$ & $ 1.797$ & $\quad 1.825(12)$ & $ 1.077$ & $ 1.016$ & $ 1.69$\\\hline
$700$ & $160$ & $ 1.210$ & $ 1.303$ & $\quad 1.328(11)$ & $ 1.098$ & $ 1.020$ & $ 1.69$\\\hline
$800$ & $120$ & $ 2.428$ & $ 2.427$ & $\quad 2.442(5)$ & $ 1.006$ & $ 1.006$ & $ 1.73$\\\hline
$800$ & $140$ & $ 1.880$ & $ 1.893$ & $\quad 1.908(5)$ & $ 1.015$ & $ 1.008$ & $ 1.72$\\\hline
$800$ & $160$ & $ 1.456$ & $ 1.479$ & $\quad 1.493(5)$ & $ 1.025$ & $ 1.009$ & $ 1.72$\\\hline
$900$ & $120$ & $ 2.290$ & $ 2.229$ & $\quad 2.237(6)$ & $ 0.977$ & $ 1.004$ & $ 1.78$\\\hline
$900$ & $140$ & $ 1.842$ & $ 1.802$ & $\quad 1.810(5)$ & $ 0.982$ & $ 1.005$ & $ 1.78$\\\hline
$900$ & $160$ & $ 1.489$ & $ 1.463$ & $\quad 1.471(3)$ & $ 0.988$ & $ 1.006$ & $ 1.78$\\\hline
$1000$ & $120$ & $ 2.087$ & $ 1.997$ & $\quad 2.002(8)$ & $ 0.959$ & $ 1.003$ & $ 1.81$\\\hline
$1000$ & $140$ & $ 1.721$ & $ 1.651$ & $\quad 1.656(7)$ & $ 0.962$ & $ 1.003$ & $ 1.82$\\\hline
$1000$ & $160$ & $ 1.429$ & $ 1.375$ & $\quad 1.380(5)$ & $ 0.966$ & $ 1.004$ & $ 1.81$\\\hline
\hline
\end{tabular}
\end{center}
{\tighten \caption{
Collection of cross sections and K factors for various c.m.\,energies and Higgs masses
and top quark mass $m_t^{\rm 1S}=180$~GeV.}
\label{tab3} }
\end{table}

In Fig.\,\ref{fig4} the Higgs energy spectrum in the large energy end point region 
is displayed at LL (dashed lines) and NLL (solid lines) order in the nonrelativistic expansion
for the vNRQCD renormalization parameters $\nu=0.1,0.2,0.4$ and for 
$\sqrt{s}=600$~GeV, $m_t^{\rm 1S}=180$~GeV, $m_H=140$~GeV, and
\begin{eqnarray}
\begin{array}{ll}
\Gamma_t=1.55~\mbox{GeV}\,, &  \\
M_Z=91.1876~\mbox{GeV}\,,   & \quad M_W=80.423~\mbox{GeV}, \\
\alpha^{-1}=137.034\,,      & \quad c_w=M_W/M_Z\,.
\end{array}
\label{parameters}
\end{eqnarray}
At LL order the upper (lower) curve corresponds to $\nu=0.1$ ($0.4$),
while at NLL order the upper (lower) curve corresponds to $\nu=0.2$
($0.1$).  The curves in Fig.\,\ref{fig4} show the typical behavior
of the nonrelativistic expansion for any choice of parameters. While
the LL predictions have a quite large renormalization parameter
dependence at the level of several tens of percent, the NLL results
are stable. Here, the renormalization parameter variation is around
5\%. The stabilization with respect to renormalization parameter
variations at NLL order arises mainly from the inclusion
of the ${\cal O}(\alpha_s)$ QCD corrections to the Coulomb potential,
Eq.\,(\ref{VCoulomb}).  
Moreover, the NLL curves lie considerably lower than the LL ones. This
behavior is well known from the predictions for $e^+e^-\to t\bar t$ at
threshold~\cite{TTBARreview,hmst} and originates from the structure of the 
large negative  ${\cal O}(\alpha_s)$ QCD corrections to the Coulomb potential,
Eq.\,(\ref{VCoulomb}), and from the sizable negative
QCD corrections to the matching conditions in
Eq.\,(\ref{matchcond}).

In principle this behavior is a point of
concern because it could indicate that the renormalization parameter
variation might be an inadequate method to estimate theoretical
uncertainties. Fortunately, the top quark mass is sufficiently large such
that the regions where the conventional fixed-order expansion (in
powers of the strong coupling) and where the
nonrelativistic expansion (described by the effective theory) can be
applied are expected to overlap.   
In Figs.~\ref{fig5} the Higgs energy spectrum is displayed in the
fixed-order expansion at the Born (dashed lines) and the 
${\cal O}(\alpha_s)$ level (solid lines) and in the nonrelativistic
expansion at LL (dotted lines) and NLL order (solid lines). For the
four panels we have chosen the parameters
(a) $\sqrt{s}=500$~GeV, $m_H=120$~GeV,
(b) $\sqrt{s}=550$~GeV, $m_H=140$~GeV,
(c) $\sqrt{s}=600$~GeV, $m_H=140$~GeV, and
(d) $\sqrt{s}=700$~GeV, $m_H=140$~GeV. For all cases we used 
$m_t^{\rm 1S}=180$~GeV as the top quark mass.
The Born curves were obtained from the analytic results given in
Ref.\,\cite{Dawson1}\footnote{
We note that the Born expression for Higgs energy distribution 
given in~\cite{Dawson2} contains typos~\cite{Dawson3}: In the
coefficient function $G_2$ a factor $s$ is missing for the term
$(-m_H^2+s+sx_h)(-\hat{\beta}^2+x_h^2)$. In the coefficient functions
$G_5$ and $G_6$ an overall factor of  minus one is missing.
}
and the ${\cal O}(\alpha_s)$ curves were obtained from the numerical
program developed by Denner et al. in Ref.\,\cite{Denner1}. 
The two ${\cal O}(\alpha_s)$ curves correspond to the renormalization
scales $\mu=\sqrt{s}$ (lower curve) and $\mu=|\sqrt{s} v|$ (upper curve),
where $v$ is the $t\bar t$ relative velocity defined in
Eq.\,(\ref{vdef}). The latter choice for the 
fixed-order renormalization scale is motivated by the fact that 
the relative momentum of the top pair is the scale that governs the
Coulomb singularities contained in the fixed-order expansion close the
large Higgs energy end point. This choice for the fixed-order
renormalization scale is therefore the more appropriate one closer to the
Higgs energy end point.
For the nonrelativistic expansion the
renormalization parameters $\nu=0.1,0.2,0.4$ were employed, as in
Fig.~\ref{fig4}. The results in Fig.~\ref{fig5}a-d demonstrate 
that there is an overlap between the ${\cal O}(\alpha_s)$ fixed-order
prediction and the NLL nonrelativistic one in the region where the
$t\bar t$ relative 
velocity is approximately 0.2. (The Higgs energy with $v=0.2$ is indicated in
each panel by the solid vertical line.) The overlap improves for increasing
c.m.\,energies or decreasing Higgs masses. This indicates that in the overlap
regions the higher order contributions summed in the nonrelativistic
prediction and the higher order relativistic corrections contained in
the fixed-order result are both small. For smaller c.m.\,energies or
increasing Higgs 
masses, on the other hand, the NLL nonrelativistic predictions tend to lie 
slightly above the ${\cal O}(\alpha_s)$ fixed-order results (for
$\mu=\sqrt{s}v$) illustrating the
impact of the higher order corrections to each type of expansion.  
The discrepancy, however, remains comparable to the uncertainties estimated
from the renormalization parameter variation of the NLL nonrelativistic
prediction. We therefore conclude that the renormalization parameter variation
of the NLL order nonrelativistic prediction should provide a realistic estimate of
the theoretical uncertainties in the large Higgs energy region.

Finally, let us discuss the numerical impact of the nonrelativistic contributions
in the large Higgs energy region on the total cross section. The
results in Figs.~\ref{fig5} show that the summation of the Coulomb
singularities and the logarithms of the top quark velocity lead to an
enhancement of the cross section and that the portion of phase space
where nonrelativistic effects are important increases for decreasing
c.m.\,energy or increasing Higgs mass. 
In Tab.\,\ref{tab3} the impact of the summations is analyzed 
numerically for various choices of the c.m.\,energy and the
Higgs mass. For all cases the top quark mass $m_t^{\rm 1S}=180$~GeV
was used.\footnote{
For the ${\cal O}(\alpha_s)$ fixed-order numbers that were obtained from the 
numerical code of Ref.~\cite{Denner1} for Fig.~\ref{fig5} and
Tab.~\ref{tab3}, we used the top pole mass as an input
with the approximation  $m_t^{\rm pole}=m_t^{\rm 1S}$. Numerically
the actual difference is of order $\alpha_s^2$, see Eq.~(\ref{mpolem1s}). 
We have checked
that this approximation is justified within the theoretical
uncertainties shown in the fifth column of Tab.~\ref{tab3}.
} 
The other parameters are fixed as in Figs.\,\ref{fig4} and
\ref{fig5} discussed earlier. In the table $\sigma(\mbox{Born})$
refers to the Born cross section and  $\sigma(\alpha_s)$ to the 
${\cal O}(\alpha_s)$ cross section in fixed-order perturbation theory
using $\mu=\sqrt{s}$ as the renormalization scale, the choice employed
in the analysis of Ref.\,\cite{Denner1}. All cross sections are given
in fb units. The term $\sigma(\mbox{NLL})$
refers to the sum of the ${\cal O}(\alpha_s)$ fixed-order cross
section for $v>0.2$ using $\mu=\sqrt{s}v$ and the NLL nonrelativistic
cross section for $|v|<0.2$ with the renormalization parameter
$\nu=0.2$. 
The numbers for $\sigma(\mbox{NLL})$ represent the currently 
most complete predictions for the total cross section of the process 
$e^+e^-\to t\bar t H$ as far as QCD corrections are concerned.  
For the results for $\sigma(\mbox{NLL})$ we have also given our
estimate for the theoretical error. For the fixed-order contribution
($v>0.2$) we have estimated the uncertainty by taking the maximum of
the shifts obtained from varying 
$\mu$ in the ranges $[\sqrt{s},2 \sqrt{s}]$, $[\sqrt{s},\sqrt{s}/2]$ 
and $[\sqrt{s}v,\sqrt{s}]$; for the nonrelativistic contribution in
the end point we have assumed an uncertainty of 5\% for all cases. 
For the errors given in Tab.~\ref{tab3} both uncertainties were added 
linearly.

The results show that the effect of the summations in the large Higgs
energy region is particularly important for smaller c.m.\,energies 
and larger Higgs masses, when the portion of the phase space where the 
nonrelativistic expansion has to be applied is large. Here, the higher
order summations contained in the nonrelativistic expansion can be
comparable to the already sizable ${\cal O}(\alpha_s)$ fixed-order
corrections and enhance the cross section further. This is advantageous
for top Yukawa coupling measurements for the lower c.m.\,energies that
are accessible in the first phase of the International Linear Collider
experiment. 
For higher c.m.\,energies the effect of the nonrelativistic summations
of contributions from beyond ${\cal O}(\alpha_s)$ is less pronounced
and decreases to the one-percent level for c.m.\,energies above $700$~GeV,
see column~seven of Tab.\,\ref{tab3}. For all cases, except for very
large c.m.\,energies around $1000$~GeV, however, the shift caused by
the terms that are summed up in the
nonrelativistic expansion exceeds the theoretical error. 
In the last column we have also displayed the ratio of the NLL
nonrelativistic cross section for $|v|<0.2$ with $\nu=0.2$ and the 
${\cal O}(\alpha_s)$ fixed-order cross section for $v<0.2$ with
$\mu=\sqrt{s}$ to illustrate the effect of the summation of QCD
corrections beyond one loop that have to be carried out in the large
Higgs energy region. Interestingly, the higher order summations lead
to correction factors ranging between about $1.7$ and $1.8$ that are only
very weakly dependent of the c.m.\,energy and the Higgs mass. 
This fact might prove useful for rough
implementations of nonrelativistic $t\bar t$ effects in other high energy 
processes.

\section{Conclusion}
\label{sectionconclusion}

In this paper we have computed the Higgs energy distribution for the
process $e^+e^-\to t\bar t H$ in the large Higgs energy end point
region. Due to Coulomb singularities $\sim (\alpha_s/v)^n$ and
logarithmic divergences $\sim (\alpha_s\ln v)^n$ that arise at higher
order of conventional multiloop-perturbation theory in the large
Higgs energy end point region, $v$ being the top velocity in the $t\bar
t$ c.m.\,frame, usual fixed-order perturbation theory breaks down. We
use the nonrelativistic effective theory vNRQCD to sum both types of
singular terms  and determine the Higgs energy distribution in the
end point region at next-to-leading logarithmic order. While the
Coulomb singularities are summed by solving the nonrelativistic
equation of motion for the $t\bar t$ Green function, the logarithmic
terms are summed using the renormalization group equations in the
effective theory. We have compared the effective theory predictions 
for the Higgs energy spectrum with the ${\cal O}(\alpha_s)$
fixed-order results from Ref.~\cite{Denner1} and can show that the
nonrelativistic and the fixed-order expansion (with the
renormalization scale $\mu$ set of order $\sqrt s\,v$) have a region
of overlap where $v$ is around $0.2$. For our next-to-leading
logarithmic order effective theory prediction we estimate a
theoretical error of 5\%. The summations lead to an additional sizable
enhancement of the total cross section compared to the positive 
${\cal  O}(\alpha_s)$ fixed-order corrections 
and are particularly important for lower c.m.~energies
(or larger Higgs masses) where the portion of the phase space where
the $t\bar t$ QCD dynamics is nonrelativistic is increased. For higher
energies the effects decrease and reach the one percent level at
c.m.\,energies around $700$~GeV. We have given updated predictions for
the total cross section $\sigma(e^+e^-\to t\bar t H)$ combining our
effective theory next-to-leading logarithmic prediction in the large
Higgs energy region with the fixed-order ${\cal O}(\alpha_s)$ results
determined in Ref.~\cite{Denner1}.

\begin{acknowledgments} 
We would like to thank S.~Dittmaier and M.~Roth for 
useful discussions and for providing us their numerical codes 
from Ref.~\cite{Denner1}.
\end{acknowledgments}

\appendix


\end{document}